\documentclass[10pt, conference, compsocconf]{IEEEtran}
\IEEEoverridecommandlockouts
\IEEEpubid{978-0-7381-4307-1/20/\$31.00 ©2020 IEEE}

\ifCLASSINFOpdf
\else
\fi
\begin{document}
%
\title{Design Methodologies for Reliable and Energy-efficient PCM Systems}


\author{\IEEEauthorblockN{Shihao Song}
\IEEEauthorblockA{Department of Electrical and Computer Engineering \\
Drexel University \\
Philadelphia, Pennsylvania, USA\\
shihao.song@drexel.edu}
\and
\IEEEauthorblockN{Anup Das}
\IEEEauthorblockA{Department of Electrical and Computer Engineering \\
Drexel University \\
Philadelphia, Pennsylvania, USA\\
anup.das@drexel.edu}
}

\maketitle

\begin{abstract}
Phase-change memory (PCM) is a scalable and low latency non-volatile memory (NVM) technology that has been proposed to serve as storage class memory (SCM), providing low access latency similar to DRAM and often approaching or exceeding the capacity of SSD. The multilevel property of PCM also enables its adoption in neuromorphic systems to build high-density synaptic storage. We investigate and describe two significant bottlenecks of a PCM system. First, writing to PCM cells incurs significantly higher latency and energy penalties than reading its content. Second, high operating voltages of PCM impacts its reliable operations. In this work, we propose methodologies to tackle the bottlenecks, improving performance, reliability, energy consumption, and sustainability for a PCM system.

\end{abstract}


%
\IEEEpeerreviewmaketitle

\section{Introduction}
Emerging non-volatile memory (NVM) technologies have been extensively studied, which is due to two reasons. 
First, there is a gap of orders of magnitude between access latencies of SSD (e.g., 0.1-1 ms) and DRAM (e.g., 10-100 ns).
Thus, there has been continuous research on finding a new layer of memory hierarchy, e.g., storage class memory (SCM)~\cite{bourzac2017has}, to fill the large performance gap. 
Second, conventional dynamic random access memory (DRAM) technology is experiencing significant scalability issues~\cite{mutlu2013memory}. Therefore, a promising NVM technology should be scalable and provide near-DRAM access latency in order to supplant or supplement DRAM for a high-performance and high-capacity main memory system.

Examples of emerging non-volatile memory technologies include phase-change memory (PCM), resistive random access memory (RRAM), ferroelectric random access memory (FeRAM), and magnetic random access memory (MRAM). This work focuses on phase-change memory (PCM), which is considered the most mature NVM technology~\cite{redaelli2018}. 


However, there are key bottlenecks in PCM.
First, PCM shows near-DRAM read latency, but its write latency is much higher than DRAM. This is because the SET operation (0 $\rightarrow$ 1 programming) requires much longer latency than RESET operation (1 $\rightarrow$ 0 programming). 
Qureshi et al. propose PreSET~\cite{QureshiISCA12}, a mechanism to proactively SET the memory location (preparing an all-ones data pattern) well in advance of the anticipated write. This enables a PCM write to perform only short-latency RESETs, improving system performance. However, RESET operations consume much more energy than SET operations. Our recent work~\cite{songISMMa} shows that PreSET can incur significant energy penalties.
Second, PCM requires high voltages and currents to operate, 
leading to accelerated aging of CMOS transistors in the peripheral circuits within each PCM bank. Each peripheral circuit consists of a sense amplifier (to read PCM cells) and a write driver (to program PCM cells). Aging of peripheral circuits impacts PCM reliability because a failed peripheral circuit cannot serve reads and writes~\cite{songISMMb,songASPDAC}.

This work addresses performance, energy consumption, and reliability issues in PCM. We make the following  key contributions.
\begin{itemize}
    \item We propose \emph{data content aware PCM writes} (\textbf{DATACON})~\cite{songISMMa}, a mechanism that lowers the latency and energy consumption of PCM writes by redirecting these requests to the best overwritten memory locations.
    \item We propose \textbf{MNEME}~\cite{songISMMb}, a novel hybrid DRAM-PCM system architecture to improve performance and reliability. MNEME consists of a segmented bitline architecture 
    and a prediction-based page management policy.
    \item We extend our study to neuromorphic hardware that uses PCM as synaptic storage elements. We propose \textbf{RENEU}~\cite{songEDCC}, a reliability-oriented approach to map machine learning applications to neuromorphic hardware, improving reliability without degrading performance.
\end{itemize}

\section{Methodologies}
\subsection{DATACON}
The key observation of PreSET~\cite{QureshiISCA12} is that overwriting an all-ones memory content incurs the least write latency because only short-latency RESETS are required. However, we observe that overwriting an all-ones memory content is not energy-efficient when the fraction of SET bits in the write data is less than 60\%, whereas overwriting an all-zeros memory content gives the best energy efficiency.

We propose \emph{data content aware PCM writes} (\textbf{DATACON}), a mechanism that lowers the latency and energy of PCM writes by redirecting these requests to memory locations that contain all-ones or all-zeros based on the number of SET bits in the write data. DATACON operates in three steps. First, for every write request, DATACON counts the number of SET bits in the write data. Second, if the fraction of SET bits in the write data is less than 60\%, DATACON will select a memory location containing all-zeros as the request's new write address. Otherwise, a memory location containing all-ones should be selected. DATACON keeps track of the address translation of the original write address to the new write address, guaranteeing the correctness for future read requests. Third, DATACON re-initializes unused memory locations, i.e., to SET or RESET one unused memory location to prepare an all-ones or all-zeros memory content. The re-initialization step employs our \emph{partition-level parallelism} (\textbf{PALP}~\cite{song2019enabling}) mechanism, avoiding interference with regular read and write requests.

We evaluate DATACON with SPEC CPU2017, NAS, and state-of-the-art machine learning workloads. Results demonstrate that DATACON improves the effective access latency by 31\%, overall system performance by 27\%, and total memory system energy consumption by 43\% compared to the best performance-oriented state-of-the-art techniques.

\subsection{MNEME}
A peripheral circuit is several orders of magnitude larger than the size of a PCM cell. To achieve higher chip density, many PCM cells share one peripheral circuit through a wire called a bitline. We observe that the farther a PCM cell from the peripheral circuit, the more parasitic components are associated with the cell. The voltage drop (called the \emph{IR drop}) induced by parasitic components need to be compensated by supplying extra voltage generated by the on-chip voltage regulators. This implies that cells nearer to their peripheral circuit can be accessed faster due to reduced latencies for voltage regulators to generate fewer voltages. Furthermore, since the lower voltage is required to access nearer cells, the aging of peripheral circuits is reduced, thus improving PCM reliability. A similar phenomenon is also observed and reported for DRAM~\cite{lee2013tiered}.

MNEME system architecture considers a computing system with PCM supplementing DRAM, forming a hybrid main memory subsystem. The first key component of MNEME is a segmented bitline architecture where each long bitline in DRAM and PCM is split into a near segment and a far segment using an isolation transistor. This new architecture creates intra-memory asymmetries for both DRAM and PCM. For instance, cells inside the near segment in PCM can be accessed faster, and the reduced voltage level also improves PCM reliability. Conversely, when accessing cells in the far segment in PCM, higher access latency and degraded reliability are expected. Thus, from the reliability and performance perspective, access-intensive pages should be placed in the near segment in PCM or DRAM.

Conventional page management policy such as Nimble \cite{yan2019nimble} does not perform well in the segmented bitline DRAM-PCM architecture. This is due to two reasons. First, Nimble does not consider intra-memory asymmetries. Second, Nimble induces significant migration overhead, which is because Nimble randomly selects a memory tier, e.g., the far segment in PCM or the near segment in DRAM, to place a new memory page during its initial allocation. Nimble then migrates pages among memory tiers based on their run-time access intensity. MNEME extends its page management policy in the following ways. First, MNEME exploits both inter-memory and intra-memory asymmetries. Second, MNEME implements a prediction-based page management policy. A new page's access-intensity is predicted then placed in a matching memory tier during its initial allocation. For instance, a page that is predicted to be write-intensive should be placed in the near segment in DRAM because of the lower operating voltage and faster access latency. This eliminates unnecessary page migrations during program execution and improves system-wide reliability as well as performance.

We evaluate MNEME with single-core and multi-programmed workloads from the SPEC CPU2017 Benchmark suites. Our results show that MNEME significantly improves the performance and reliability of state-of-the-art hybrid memory systems. 

\subsection{RENEU}
Neuromorphic hardware is designed for executing spiking neural networks (SNN). Recently, non-volatile memory technologies such as phase-change memory are used to implement synaptic storage in neuromorphic hardware~\cite{songLCTES}. PCM brings certain advantages such as multilevel storage and high integration density. 

We analyze the internal circuitry of a (pre-synaptic) neuron and observe that due to PCM's high operating voltages, the CMOS transistors inside the neuron are stressed at elevated voltages when propagating excitations through PCM synapse, leading to accelerated CMOS aging. This impacts the lifetime reliability of the neuromorphic hardware. In our study, we concentrate CMOS aging considering Time-Dependent Dielectric Breakdown (TDDB), Negative-Bias Temperature Instability (NBTI), and Hot-Carrier Injection (HCI) failure mechanisms. These are the dominant failure mechanisms in 45nm and below technology nodes~\cite{balajiCAL,songMWCAS}.

Our key observation is that CMOS aging in a neuron circuitry is dependent on the number of spikes/excitations propagated by the neuron. Thus, we formulate and express a single neuron's failure mechanisms in terms of the number of spikes propagated by it. We consider a tile-based neuromorphic hardware and extend the formulation to incorporate all the neurons in a tile. To this end, we consider a tile is faulty if any of the neurons in it fails. Similarly, a single faulty tile leads to the failure of the neuromorphic hardware. Formulating failure mechanisms as a series failure model enables us to apply mapping techniques to minimize the maximum aging, e.g., a mapping technique that balances the aging among all the tiles.

We propose \textbf{RENEU} (\underline{RE}liability-aware \underline{NEU}romorphic Computing), a reliability-oriented approach to map spiking neural networks to neuromorphic hardware, improving system-side reliability without compromising key performance metrics such as execution time. RENEU first partitions an SNN model into clusters, where each cluster with a fixed number of neurons and synapses can fit its entirety to a tile. Incorporating the reliability formulation, RENEU then uses the clustered SNN model to find the cluster-to-tile mapping based on Particle Swarm Optimization (PSO). 

We evaluate RENEU using different machine learning applications on a state-of-the-art neuromorphic hardware. Results demonstrate an average 38\% reduction in circuit aging, leading to an average 18\% improvement in the hardware's lifetime compared to current practices. RENEU only introduces a marginal performance overhead of 5\% compared to our performance-oriented mapping mechanism~\cite{balaji20pycarl}.

\section{Conclusion}
We introduce DATACON and MNEME to address performance, energy consumption, and reliability issues of conventional PCM-based computing systems. We also extend our study on PCM to neuromorphic hardware and propose RENEU to improve its lifetime reliability. We evaluate our methodologies using state-of-the-art workloads, and results show that they significantly improve performance, energy consumption, and reliability for PCM-enabled systems. In future research, we plan to extend our study on PCM to the following potential fields.
\begin{enumerate}
    \item \textbf{GPU-PCM architecture}. GPUs are facing challenges such as limited on-board DRAM space. We plan to explore the performance and power bottleneck when supplanting or supplementing DRAM with PCM. 
    \item \textbf{Managed programming languages and in-PCM computing}. Recent research shows that the garbage collection mechanism in managed programming languages such as Java induce significant performance penalties. 
    We plan to explore in-memory computing techniques to mitigate performance bottleneck for managed programming languages executed on PCM-based computing systems.
\end{enumerate}


\section*{Acknowledgment}
This work is supported by 1) the National Science Foundation Faculty Early Career Development Award CCF-1942697 (CAREER: Facilitating Dependable Neuromorphic Computing: Vision, Architecture, and Impact on Programmability) and 2) the National Science Foundation Award CCF-1937419 (RTML: Small: Design of System Software to Facilitate Real-Time Neuromorphic Computing).



%

\bibliographystyle{IEEEtran}
\bibliography{references}

\begin{thebibliography}{10}
\providecommand{\url}[1]{#1}
\csname url@samestyle\endcsname
\providecommand{\newblock}{\relax}
\providecommand{\bibinfo}[2]{#2}
\providecommand{\BIBentrySTDinterwordspacing}{\spaceskip=0pt\relax}
\providecommand{\BIBentryALTinterwordstretchfactor}{4}
\providecommand{\BIBentryALTinterwordspacing}{\spaceskip=\fontdimen2\font plus
\BIBentryALTinterwordstretchfactor\fontdimen3\font minus
  \fontdimen4\font\relax}
\providecommand{\BIBforeignlanguage}[2]{{%
\expandafter\ifx\csname l@#1\endcsname\relax
\typeout{** WARNING: IEEEtran.bst: No hyphenation pattern has been}%
\typeout{** loaded for the language `#1'. Using the pattern for}%
\typeout{** the default language instead.}%
\else
\language=\csname l@#1\endcsname
\fi
#2}}
\providecommand{\BIBdecl}{\relax}
\BIBdecl

\bibitem{bourzac2017has}
K.~Bourzac, ``{Has Intel created a universal memory technology?}'' \emph{IEEE
  Spectrum}, 2017.

\bibitem{mutlu2013memory}
O.~Mutlu, ``Memory scaling: A systems architecture perspective,'' in
  \emph{IMW}, 2013.

\bibitem{redaelli2018}
A.~Redaelli \emph{et~al.}, ``{Phase change memory: device physics, reliability
  and applications},'' \emph{Springer}, 2018.

\bibitem{QureshiISCA12}
M.~K. Qureshi \emph{et~al.}, ``{PreSET}: Improving performance of phase change
  memories by exploiting asymmetry in write times,'' in \emph{ISCA}, 2012.

\bibitem{songISMMa}
S.~Song, A.~Das, O.~Mutlu, and N.~Kandasamy, ``Improving phase change memory
  performance with data content aware access,'' in \emph{ISMM}, 2020.

\bibitem{songISMMb}
S.~Song, A.~Das, and N.~Kandasamy, ``Exploiting inter- and intra-memory
  asymmetries for data mapping in hybrid tiered-memories,'' in \emph{ISMM},
  2020.

\bibitem{songASPDAC}
S.~Song, A.~Das, O.~Mutlu, and N.~Kandasamy, ``Aging aware request scheduling
  for non-volatile main memory,'' in \emph{ASP-DAC}, 2021.

\bibitem{songEDCC}
S.~Song, A.~Das, and N.~Kandasamy, ``Improving dependability of neuromorphic
  computing with non-volatile memory,'' in \emph{EDCC}, 2020.

\bibitem{song2019enabling}
S.~Song, A.~Das, O.~Mutlu, and N.~Kandasamy, ``Enabling and exploiting
  partition-level parallelism ({PALP}) in phase change memories,'' \emph{TECS},
  2019.

\bibitem{lee2013tiered}
D.~Lee, Y.~Kim, V.~Seshadri, J.~Liu, L.~Subramanian, and O.~Mutlu,
  ``{Tiered-latency DRAM: A low latency and low cost DRAM architecture},'' in
  \emph{HPCA}, 2013.

\bibitem{yan2019nimble}
Z.~Yan, D.~Lustig, D.~Nellans, and A.~Bhattacharjee, ``Nimble page management
  for tiered memory systems,'' in \emph{ASPLOS}, 2019.

\bibitem{songLCTES}
S.~Song, A.~Balaji, A.~Das, N.~Kandasamy, and J.~Shackleford, ``Compiling
  spiking neural networks to neuromorphic hardware,'' in \emph{LCTES}, 2020.

\bibitem{balajiCAL}
A.~Balaji, S.~Song, A.~Das, N.~Dutt, J.~Krichmar, N.~Kandasamy, and
  F.~Catthoor, ``A framework to explore workload-specific performance and
  lifetime trade-offs in neuromorphic computing,'' \emph{CAL}, 2019.

\bibitem{songMWCAS}
S.~Song and A.~Das, ``A case for lifetime reliability-aware neuromorphic
  computing,'' in \emph{MWCAS}, 2020.

\bibitem{balaji20pycarl}
A.~Balaji, P.~Adiraju, H.~J. Kashyap, A.~Das, J.~L. Krichmar, N.~D. Dutt, and
  F.~Catthoor, ``{PyCARL}: A pynn interface for hardware-software co-simulation
  of spiking neural network,'' in \emph{IJCNN}, 2020.

\end{thebibliography}




\end{document}